\documentclass[]{sig-alternate}

\usepackage[T1]{fontenc} 
\usepackage[utf8]{inputenc} 
\usepackage[]{algorithm2e}
\usepackage{multirow}
\usepackage{xcolor}
\usepackage{balance}
\usepackage{xspace}
\usepackage{listings}
\usepackage{comment}

\newcommand{\nopol}{{\sc Nopol}\xspace}
\def\ourif{{\sc if}\xspace}

\title{Automatic Repair of Buggy If Conditions and Missing Preconditions with SMT}

\numberofauthors{4}
\author{
\alignauthor Favio DeMarco\\
       \affaddr{Universidad de Buenos Aires}\\
       \affaddr{Buenos Aires, Argentina}\\
\alignauthor Jifeng Xuan\\
       \affaddr{INRIA Lille - Nord Europe}\\
       \affaddr{Lille, France}\\
\and
\alignauthor Daniel Le Berre\\
       \affaddr{University of Artois \& CNRS}\\
       \affaddr{Lens, France}\\
\alignauthor Martin Monperrus\\
       \affaddr{University of Lille \& INRIA}\\
       \affaddr{Lille, France}\\       
}

\begin{document}
\conferenceinfo{CSTVA}{'14, May 31 – June 7, 2014, Hyderabad, India}
\CopyrightYear{14}
\crdata{978-1-4503-2847-0/14/05}
\maketitle

\begin{abstract}
We present \nopol\, an approach for automatically repairing buggy \ourif conditions and missing preconditions.
As input, it takes a program and a test suite which contains passing test cases modeling the expected behavior of the program and at least one failing test case embodying the bug to be repaired. 
It consists of collecting data from multiple instrumented test suite executions, transforming this data into a Satisfiability Modulo Theory (SMT) problem, and translating the SMT result -- if there exists one -- into a source code patch.
\nopol\ repairs object oriented code and allows the patches to contain nullness checks as well as specific method calls.
\end{abstract}




\section{Introduction}
Automatic software repair consists in automatically fixing known bugs in a program. For instance, an automatic software repair approach can generate a patch that makes a failing test case pass. 
This is ``test-suite based program repair'', as pioneered by Le~Goues et al. \cite{Goues2012}, and further explored by Nguyen et al. \cite{nguyen13} as well as Kim et al. \cite{Kim2013}.
The main motivation of automatic software repair is to decrease the cost of fixing bugs. 
The synthesized patches can be proposed as potential solutions to developers \cite{weimer2006patches} or used as is when it is urgent. In the latter case, it is believed that having a draft solution eases the time needed to comprehend the bug and to design a fix \cite{Goues2012,nguyen13,Kim2013}.

In the context of automatic repair, a fault model refers to the kind of bugs that can be fixed with a given approach \cite{koopman2003elements,monperrus14}.
In this paper, we concentrate on the following fault model: the automatic repair of buggy \ourif conditions and missing preconditions.
Both are members of the family of condition-related bugs.
Pan et al. \cite{pan2009toward} as well as Martinez and Monperrus \cite{Martinez2013} have shown that fixes of such bugs are among the most common ones.
Our repair approach, called \nopol\footnote{\nopol\ is an abbreviation for ``No-Polillas'' in Spanish, which literally means ``No-Moth''.}, repairs buggy \ourif conditions and missing precondtions of object-oriented source code written in Java. 

For instance, \nopol\ can synthesize a patch that adds a precondition as shown in Listing \ref{example-patch}.

\begin{lstlisting}[columns=fullflexible,float,label=example-patch,captionpos=b,caption={Example of synthesized patch. \nopol\ repairs buggy \ourif conditions and missing precondtions of object-oriented source code written in Java.}]
+if (l!=null && l.size()>0) {
  compute(l);
+}
\end{lstlisting}

As previous work on repair,  \nopol\ is test-suite based. As input, it takes a program and its test suite, given that there is at least one failing test case embodying the bug to be repaired. 
Then, \nopol\ analyzes one by one the statements that are executed by the failing test case in order to identify those for which there may exist a patch. This is what we call ``fix point localization'': finding potential fixing locations (as opposed to fault localization which consists of finding potential root causes). 

In the process of finding potential fixing locations, \nopol\ identifies fix oracles. In the example above, that means we know that skipping statement \texttt{compute()} would enable the test to pass.
Of course, completely removing the statement is not an option, it would break the passing test cases.
In other words, we know that the synthesized precondition must output ``false'' for the failing test case\footnote{Also, ``true'' for the passing ones.}. The ``false'' is a fix oracle.

Then, \nopol\ collects information from test suite execution through code instrumentation. This information is basically the local program state at each fix point. This information contains both ``primitive'' values (integer, booleans) as well as object-oriented ones (nullness, object state). 
Next, those runtime traces are transformed into a Satisfiability Modulo Theory (SMT) problem. An SMT solver says whether there exists a solution. 
If such a solution exists, it is translated back as source code, i.e. a source code patch is generated.

\nopol\ provides a complete approach for fixing buggy \ourif conditions and missing precondtions. It blends existing techniques with novel ideas.
From the literature, we reuse:
the idea of fixing buggy \ourif conditions \cite{nguyen13},
the concept of artificially manipulating the program state during program execution \cite{zhang2006locating,jeffrey2008fault},
the encoding of program synthesis as an SMT problem \cite{gulwani2011synthesis,jha2010oracle}.
The novelty of this paper lies in:
\begin{itemize} 
\itemsep0em 
\item the design of a repair approach for a new kind of fault: missing preconditions;
\item  an algorithm called ``angelic fix localization'' for identifying at once potential repair locations and repair oracles;
\item the extension of the SMT encoding for handling nullness and certain methods calls of object-oriented programs;
\item a case study of a real bug on a real-world large piece of software, the Apache Common Maths library: 5000 lines of executable code and 352 test cases (test methods in JUnit).
\end{itemize}

\section{Background}
\label{background}

\subsection{Test Suite based Program Repair}
\label{sec:test-suite-based-repair}

Test-suite based program repair consists in repairing programs for which a test suite is available
and for which at least one failing test case states the bug.
This failing test case is either a regression bug or a new bug that has just been discovered.
Then, repair algorithms search for patches that make the failing test pass while keeping the other test cases green.
If such a patch is found, the patch fixes the bug and at the same time does not degrade the existing functionalities.
Test-suite based program repair has been mostly disseminated by the work of Le~Goues et al. on GenProg \cite{Goues2012}, and is an actively explored research area \cite{nguyen13,Kim2013}.

\subsection{Oracle-based Program Repair}
\label{sec:oracle-based-repair}

Nguyen et al. \cite{nguyen13} have invented the idea of oracle-based program repair. 
It decomposes program repair in two phases.

First, oracle-based program repair looks for a pair  $(l, v)$ that would fix the bug under repair as follows.
If one uses the value $v$ at source code location $l$, the bug is fixed.
Nguyen et al. defines two kinds of pairs $(l, v)$:
(arithmetic assignment, arithmetic value) and (\ourif conditions, boolean value).
The former could state for instance ``if one assigns $3$ to variable $x$ at line 42 of Foo.java, the bug is fixed''. The latter expresses ``if the \ourif condition at line 57 of Foo.java is falsified, the bug is fixed".
In Nguyen's terms, the value $v$ ($3$ and $false$ in our examples) is called the oracle.
We call this phase \emph{``oracle mining''}.

The second phase consists in finding a piece of code that, given the program state at location $l$ would yield value $v$. The patch is correct if and only if for all executions of location $l$ (not only the buggy one), the synthesized expression outputs a correct value.  We call this phase \emph{''repair synthesis'}.

In the context of test-suite program repair, repair synthesis means for all executions of location $l$ by test cases, the synthesized expression must output a value that eventually enables the test case to pass. Note that a location can be executed $n$ times in the same test case. In this case, the synthesized expression must output $n$ values that, given all combinations and effects of the program, yield a passsing test case. 

For oracle mining, Nguyen et al. \cite{nguyen13} use symbolic execution: the oracle is a value that satisfies a set of constraints. For repair synthesis, they use oracle-guided program synthesis \cite{jha2010oracle}. 
However, both phases can be implemented in different ways. For instance, in this paper, we replace symbolic execution with another technique (see Section \ref{sec:angelic-fix-localization}). Repair synthesis can be based on constraint satisfaction or on evolutionary computation \cite{Goues2012}.

An interesting property of oracle-based program repair is to replace the concept of ``fault localization'' by the concept of ``repair localization''.
While ``fault localization'' emphasizes finding the root cause (the fault) to fix it, ``repair localization'' emphasizes finding places where a fix can be written. 
For instance, it may happen that there exists several different variables for which changing the initialization value fixes the bug. 
``Repair localization" is more pragmatic than fault localization: in real-world life there are many cases where the root cause can not be identified at a reasonable cost, but where it is sufficient to mitigate the error propagation. Note that one of the repair locations can indeed be the root cause of the fault.

Each pair $(location, value)$ is a potential fix locations. If a patch can be correctly synthesized for all executions, it becomes an actual fix location.
Indeed, there may be different locations for which there exists a value that fixes the bug.

This corresponds to what developers know in their every day bug fixing activities:
the very same bug can be fixed at different places.
In particular, it occurs that a bug can be fixed at a place that is different from the root cause. 
In this case, one does not prevent the bug to appear, but one prevents the fault to be propagated.

\subsection{Buggy \ourif condition bugs}

Conditional statements (e.g., if (condition) \{...\} in Java), are widely-used in programming languages. Pan et al. \cite{pan2009toward} show that among seven studied Java projects, up to 18.6\% of bug fixes have changed a buggy condition in if statements. For a buggy program, a buggy \ourif condition may lead to a different branch. In this paper, our tool, \nopol, is motivated by addressing fixing conditional bugs. A buggy \ourif condition is defined as a bug in the condition of an if/then/else statement. Pan et al. \cite{pan2009toward} divide buggy \ourif condition fixing into six sub-patterns, i.e., the addition/removal of a clause, a variable, an operator.   
The following example is a real example of buggy \ourif condition found in Apache Commons Math library. $gcd$ is a method of calculating the greatest common divisor between two integers. A condition in that method is to check if any of the two parameters $u$ and $v$ is equal to 0. In the buggy version, the developer compares the product of the two integers to zero. However, this may lead to an arithmetic overflow. A safer way to proceed is to compare each parameter to zero.
Such fix is synthesizable by \nopol.
\begin{verbatim}
   public static int gcd(int u, int v) {
-      if (u * v == 0) {
+     if ((u == 0) || (v == 0)) {
\end{verbatim}

We will explain how to fix buggy if condition bugs in Section \ref{sec:contribution} and two case studies of solving this kind of bugs can found in Sections \ref{case:semfix} and \ref{sec:bug-math}.  

\subsection{Missing precondition bugs}
Another type of common bugs related to branching is missing preconditions. One of the usages of preconditions is to distinguish different values of a variable, e.g., detecting null pointers or invalid indexes in an array. Developers add preconditions to ensure the program meets their expectation of variables. We define a missing precondition as a bug without its proper preconditions. 

An example of missing preconditions is the absence of null-pointer detection as follows. The buggy version without if will throw an exception signaling a null-pointer at runtime. A case study of solving this kind of bugs can found in Section \ref{sec:bug-misspre}.  
\begin{verbatim}
+  if (directory != null) 
           File[] files = directory.listFiles();
\end{verbatim}

\section{Our approach}
\label{sec:contribution}
This section presents our approach for automatically repairing buggy \ourif conditions and missing preconditions in Java source code.  
Our approach blends existing ideas (such as encoding the program synthesis in SMT) with new ones (such as angelic fix localization).
This approach has been implemented in a tool called \nopol.

\subsection{Overview}
\nopol\ is a  test-suite based program repair approach dedicated to incorrect \ourif conditions and missing precondition bugs. 
\nopol\ requires a test suite which represent the program expected functionality in which a failing test case represents the bug to be fixed. 
\nopol\ uses a novel technique to identify potential repair locations (angelic localization).
For each repair location, \nopol\ runs the whole test suite in order to collect the context
of the conditional expression and its expected value in each test.
Then such information is used to generate an SMT formula modeling expressions which
preserves the behavior of the expression for passing tests while modifying it for failing tests.
If this SMT formula is satisfiable, the SMT solution is translated as a source code patch.
\nopol\ supports a subset of object-oriented primitives (nullness and certain method calls).
For instance, \nopol\ can output:
\begin{verbatim}
Fix found!
At line 348 of file Foo.java, replace
  if (a<b)
by
  if (l!=null && l.size()<0)
\end{verbatim}

\subsection{Angelic fix localization}
\label{sec:angelic-fix-localization}

As presented in Section \ref{sec:oracle-based-repair}, oracle-based program repair needs pairs $\{location, value\}$. SemFix \cite{nguyen13} uses symbolic execution for extracting those pairs.
In \nopol, we propose to use value replacement \cite{jeffrey2008fault}  for this.

Value replacement \cite{jeffrey2008fault} comes from fault localization research. It consists in replacing at runtime one value by another one. 
More generally, the idea is to artificially change the program state for locating faults. 
There are a couple of papers that explore this idea. 
For instance, Zhang et al. \cite{zhang2006locating} calls it ``predicate switching'' and Chandra et al. \cite{chandra2011angelic} use the term ``angelic debugging''. 

In this paper, we call \emph{``angelic fix localization''} the technique of modifying the program state 
to find angelic pairs $(l, v)$. 

{\bf Definition (angelic value)}
An angelic value is a value that is arbitrarily set during test execution by an omniscient angel and that enables a test to pass.

Angelic fix localization gives two key pieces of information:
first, where a fix point may exist (a fix consists of a piece of code put at a certain fix point),
second the expected value for driving the repair synthesis.
Our key insight is that the search space of angelic values is small for two kinds of bugs and locations:
the buggy \ourif conditions and the missing preconditions.

\subsubsection{For Buggy \ourif Conditions}

\begin{algorithm}[t] 
 \KwData{\\- $F$ is a failing test case\\
- $Trace$ is the execution trace of F
}
 \KwResult{$R$ is a set of pairs (location, angelic value)}
$R \leftarrow \emptyset$ \;
 \ForEach{\ourif condition $i$ in Trace}{
  force $i$ to $true$\\
  \eIf{test case F passes}{
  $R \leftarrow R \cup  \{(i, true)\}$
   }{
   force $i$ to $false$\\
   \If{test case $F$ passes}{
   $R \leftarrow R \cup \{(i, false)\}$
   }
  }
 }
\Return{$R$}
 \caption{Angelic Fix Localization Algorithm for Buggy \ourif Conditions}
\label{alg:if-angelic-fix-localization}
\end{algorithm}

For buggy \ourif conditions, angelic fix localization works as follows.
For each \ourif condition that is evaluated during the test suite execution, an angel forces the \ourif condition to be evaluated to true or false in the failing test case embodying the bug to be fixed. 
For a pair (\ourif condition, boolean value), if the failing test case now passes, it means that we have found 
a potential fix location and an oracle for SMT program repair.
Algorithm \ref{alg:if-angelic-fix-localization} is  the pseudocode of this algorithm.
One really sees the parallel with angelic debugging: when the failing test case executes, an angel comes and forces the \ourif condition to be evaluated to either true or false, i.e. forces the program to take one or the other branch.

It may happen that the same \ourif condition is executed several times in the same test case. In this case, angelic fix localization considers the expression always yield the same value. Doing so, the search space of angelic fix localization algorithm for buggy \ourif conditions is small.
Let $n$ be the number of \ourif conditions that are executed in the failing test case.
The search space is simply $2\times n$ (checking independently $n$ binary values). In practice, according to our experience with open-source test suites, a test case executes around $10^1$--$10^2$ ifs and rarely more than $10^3$.
Note that our angelic fix localization algorithm for buggy \ourif conditions only requires to run the failing test case, not the entire test suite.

\subsubsection{For Missing Preconditions}
The angelic fix localization for missing preconditions is slightly different from angelic fix localization for \ourif conditions.
For each statement\footnote{It works also for missing precondition for blocks since in Java blocks are just specific statements.} that is evaluated during the test suite execution, an angel forces to skip it.
If the failing test case now passes, it means that a potential fix location has been found.
The oracle for repair is then  ``False'', meaning that the precondition to be synthesized must output False  (i.e. the statement should be skipped).
Algorithm \ref{alg:missing-pred-angelic-fix-localization} is the pseudocode of this algorithm.
Again, it is only a potential fix location. The repair synthesis may fail to find an expression which evaluates to false in all test cases but the failing one.

Similarly to  angelic fix localization for buggy \ourif conditions, if a statement is executed several times in the same test case,  angelic fix localization completely skips it.
The size of the search space of angelic fix localization for missing preconditions is simply the number of executed statements. For each of them, the failing test case is run once with one statement skipped.

\subsubsection{Discussion}

\nopol\ uses an existing fault localization technique to increase the likelihood of finding an oracle.
The statements are not skipped randomly, but according to their ``suspiciousness". The suspiciousness of a statement measures its likelihood of containing a fault. \nopol\ uses the Ochiai spectrum based metric \cite{abreu2006evaluation} for that purpose. Given a program and a test suite, the suspiciousness $susp(s)$ of a statement $s$ is defined as follows. 
\begin{align*}
susp(s) =  \frac{failed(s)}{\sqrt{total\_failed*(failed(s)+passed(s))}}
\end{align*}
where $total\_failed$ denotes the number of all the failing test cases and $failed(s)$ and $passed(s)$ respectively denote the number of failing test cases and the number of passing test cases, which cover the statement $s$. \ourif conditions are also evaluated with the angel based on suspiciousness. We rank all the \ourif conditions based on their suspiciousness. The angel first manipulate the most suspicious executed if (resp. statement), then the second one, etc.

If no angelic pair can be found, it may mean two things.
First, if the \ourif condition (resp. statement) is executed only once in the failing test case, it means that we know for sure that it is impossible to fix the bug by changing this particular condition (resp. adding a precondition before this statement).
Second, if the \ourif condition (resp. statement) is executed more than once in the failing test case (say $p$ times), there may exist a sequence of $p$ angelic values (say true, true, false, true) resulting in a passing test case.
However, recall that \nopol assumes that, for a given test case, the angelic value is the same during the whole test case execution (for sake of having a small search space -- $2$ instead of  $2^p$ see \ref{sec:angelic-fix-localization}).
Consequently, angelic sequences are not taken into account (finding of those sequences is computationally too expensive when $n$ is large).
How does this affect the effectiveness of the tool? According to our experiments with real test suites, most \ourif conditions are evaluated only once per test case. A systematic empirical study on this point is future work.

SemFix uses symbolic execution for oracle mining and symbolic execution is known to be heavyweight. In their paper \cite{nguyen13}, they apply it only on small examples. We have applied it to real bugs, one being presented in Section \ref{sec:evaluation}. We do not have access to the code of SemFix, so we cannot measure their execution time against ours.

\begin{algorithm}[t] 
 \KwData{
\\- $F$ is a failing test case\\
- $Trace$ is the execution trace of $F$
}
 \KwResult{$R$ is a set of pairs (location, angelic value)}
$R \leftarrow \emptyset$ \;
 \ForEach{statement $s$ in $Trace$}{
  force $s$ to be skipped\\
  \If{test case $F$ passes}{
  //the precondition should evaluate to false\\
  // to skip $s$ \;
    $R \leftarrow R \cup \{(s, false)\}$
   }
 }
 \caption{The Angelic Fix Localization Algorithm for Missing Preconditions}
\label{alg:missing-pred-angelic-fix-localization}
\end{algorithm}

\subsection{Runtime Trace Collection for Repair}
\label{sec:data-collection} 
Once \nopol has found an angelic pair (location, value), it collects the values that are accesible at this point in the program execution. 
Those value are meant to be used to synthesize a correct patch.
There are different kinds of data to be collected.

\subsubsection{Primitive Type Data Collection}

At the location of an angelic pair, \nopol\ collects the values of all local variables, method parameters and fields that are typed with a basic primitive type (integer, float, boolean, etc.).
They form the core of $C_{l,m,n}$ the set of collected values at location $l$ during the  $m$-th execution of the $n$-th test case.
$C_{l,m,n}$ is also enriched with constants for further use during synthesis.

In order to be able to synthesize conditions that use literals (e.g. ``if (x>0)''), we create a set of predefined constants to be passed to the SMT solver afterwards. We have set so far two strategies for creating the set of predefined constants. The first one consists of \{0, -1, 1\}, it is a baseline that fixes many conditional bugs related to emptiness and off-by-one. The second one consists of collecting all numerical literals of the codebase. Assessing the effectiveness  of both strategies is out of the scope of this paper.

\subsubsection{Expected Outcome Data Collection}

Let us call $O$ the set of expected outcomes in order to pass all tests. 
$O_{l,m,n}$ is the expected outcome at location $l$ during the  $m$-th execution in order to pass the  $n$-th test case.

\emph{For buggy \ourif conditions.} $O_{l,m,n}$ is the expected outcome of the condition expression $l$.
For failing test cases, the expected outcome is the angelic value. 
For passing test cases, the expected outcome is the actual one, i.e. the result of the evaluation of the actual \ourif condition expression.

\begin{align*}
O_{l,m,n} = & \left\{ \begin{array}{l}
        eval(l) \mbox{~for passing test cases} \\
        \mbox{angelic value for failing test cases} \\
    \end{array}\right. \\
\end{align*}

\emph{For missing preconditions.} $O_{l,m,n}$ is the expected value of the precondition of statement $l$, i.e. true if all cases but the failing test cases. The latter comes from angelic fix localization: if the precondition returns false for the failing test case, the buggy statement is skipped and the test case passes.

\begin{align*}
O_{l,m,n} = & \left\{ \begin{array}{l}
        true \mbox{~for passing test cases} \\
        false \mbox{~for failing test cases} \\
    \end{array}\right. \\
\end{align*}

\nopol\ collects $O_{l,m,n}$ for all execution of location $l$.

\subsubsection{Object-oriented Specific Data Collection}

\nopol\ aims at supports the automatic repair of \ourif conditions and missing preconditions of object-oriented programs. 
In particular, we would like to support not-null checks and method calls to some extent.
For instance, we would like to be able to synthesize the following missing precondition.
\begin{verbatim}
+if (l!=null && l.size()>0) {
  compute(l);
+}
\end{verbatim}

To do so, in addition to collecting all values of primitives types, \nopol collects two kinds of information.
First, the nullness of object encodes whether all objects of the current scope are null or not.
Second, \nopol collects the output of ``state query methods'' as defined as the methods that enable one to inspect the state of objects and are side-effect free.
For instance, methods \emph{size()} and \emph{isEmpty()} on collections are state query methods.

\nopol\ is manually fed with a list of such methods. The list is set with domain-specific knowledge. For instance, in Java, it is easy for developers to identify such side effect free state query methods on core library classes such as String, File and Collections. For each type $T$, those predefined methods are denoted $sqm(T)$.

\nopol\ collects the nullness and the evaluation of state query methods for all objects in the scope (local variable, method parameter, fields) of an angelic pair.

\subsubsection{Repair  Equation}

The repair synthesis of buggy \ourif conditions and missing preconditions consists in finding an expression (a function) $exp$ such that
\begin{equation}
\forall_{l,m,n} \mbox{~~}exp(C_{l,m,n}) =  O_{l,m,n} 
\label{eq:repair-if}
\end{equation}

\subsubsection{Number of Collected Values}

Let us assume there are $j$ primitives values and $k$ objects (denoted $O$) in the scope of an angelic pair. 
In total, \nopol\ collects the following values:
\begin{itemize}
\item the $j$ primitive variables in the scope;
\item the $k$ boolean values corresponding to the nullness of each object;
\item $\sum_{o \in O}|sqm(type(o))|$ values corresponding to the evaluation of the state query methods of all objects available in the scope;
\item the constants.
\end{itemize}
All this information is used for finding a solution statisfying Equation~\ref{eq:repair-if}.
There are different ways of finding such as solution. 
\nopol, as SemFix \cite{nguyen13}, uses a variation of oracle-guided component-based program sythesis \cite{jha2010oracle} based on SMT.

\subsection{Encoding Repair in SMT}

We now present how we encode Equation~\ref{eq:repair-if} as an SMT problem.
The solution of the SMT problem is then translated back as a boolean source code expression $exp$ representing the correct \ourif conditional or the missing precondition.
Our encoding extends the SMT encoding defined in \cite{nguyen13,jha2010oracle}.
In particular, we explicitly take into account the type of the variables so that a boolean expression can mix operations on booleans, integers and real.

\subsubsection{Building Blocks}
\label{sec:bb}

We define a {\em building block} (called {\em component} in \cite{jha2010oracle}) as a type of expression that can appear in the boolean expression to be synthesized. 
For instance, the logical comparison operator ``>'' is a building block. As building block types, we consider  comparison operators ($>$,$<$, $\neq$, $\leq$, $=$, $\geq$), the three arithmetic operators\footnote{Adding the division is possible but would require specific care to avoid division by zero.} ($+, -, \times$) and the boolean operators ($\wedge$, $\vee$, $\lnot$). The same type of building blocks can appear multiple times in the same expression.

We define the $i$th building block $b_i$ as a tuple of input variables $I_i$, an output variable $r_i$, an expression $\phi_{i}(I_i,r_i)$ encoding the meaning of the building block (e.g.  $r_i = \wedge I_i$). That is $b=(\phi_{i}(I_i,r_i),I_i,r_i)$.
$r$ is the return value of the synthesized expression, hence there exists one building block $i$ whose output is bound to the return value $r_i = r_{final}$.

Suppose we are given a set $B$ of building blocks and a list $CO$ of pairs $(C_{l,m,n},O_{l,m,n})$ (the collected values and angelic oracles at location $l$ during the  $m$-th execution of the $n$-th test case). $C_{l,m,n}$ includes values of different type: boolean, integer or real expressions. A patch is a sequence of building blocks $<b_1,b_2,...,b_k>$ with $b_i \in B$, whose input values are either taken from $C_{l,m,n}$ or from other building blocks.

\subsubsection{Wiring}
The problem is thus to wire the input of the building blocks $<b_1,b_2,...,b_k>$ to the input values of the program $I_0$ or to other building block's output values and to make sure that one building block produces the expected output value $r$ (the angelic value). Compared to previous work \cite{jha2010oracle,nguyen13}, we need to make sure that the type of the variables are valid operands (i.e. that an arithmetic operator only manipulates integers, etc.). 

Let us assume $C_{l,m,n}$ having two values ``False'' (boolean) and ``3'' (integer), and two building blocks:\\ $BOOL \leftarrow f_1(BOOL)$ and 
$BOOL \leftarrow f_2(INT, INT)$.\\
The synthesis consists of finding a well formed expression combining $False$, $f_1(False)$, $f_2(3,3)$, $f_1(f2(3,3))$ which evaluates to $r$ the angelic value. 

\subsubsection{Mapping Inputs and Outputs with Location Variables}

We define $I = \cup I_i$ and $O = \cup \{r_i\}$ the sets of input and output values of the building blocks. 
Let $I_0$ be the input ($C_{l,m,n}$) and $r$ the output in the final patch.
We define $IO$ as $IO = I \cup O \cup I_0 \cup \{ r \}$. 
We partition the variables of $IO$ according to their type in BOOL, INT and REAL.

The SMT encoding relies on the creation of {\em location variables} $L = \{ l_x | x \in IO \}$ representing an index of elements $x\in IO$. {\em Value variables} $V = \{v_x | x \in IO \}$ represent the values taken by those elements. Let $m$ be the number of possible inputs $m=|I_0|+|B|$. Location variables are of type integer ($L \subset INT$).
Value variables are of any supported type (boolean, integer or real). 
Location variables are invariants for all test case execution $C_{l,m,n}$: they represent the patch structure.
Value variables are used internally by the SMT solver to ensure that the  semantic of the
program is preserved.

\subsubsection{Constraints}

Let us first define the domain constraints over the location variables.
The location variables of the elements of $I_0$ and $r$ are fixed:

$$\phi_{FIXED}(I_0,r) = \bigwedge_{i=1}^{|I_0|} l_{I_0,i} = i \land l_r = m$$

The location variables of the elements of $O$ have a domain of $[|I_0|+1, |I_0|+m]$:
$$\phi_{OUTPUT}(B) = \bigwedge_{i=1}^{|B|} ( |I_0|+1 < l_{b_i} \leq m)$$

\textbf{Handling typing} Only the locations corresponding to the values of the same type are allowed.
Suppose that $type(x)$ returns the set of elements with the same type than $x$ among BOOL, INT and REAL. Then we can restrict the values taken by the location variables of the input values of building blocks using the following formula:

$$\phi_{INPUT}(I) = \bigwedge_{x \in I} \bigvee_{y \in type(x), x \neq y} (l_x = l_y)$$

In our example, for a single input value of type integer, we have the following domains for location variables:
\begin{description}
\setlength{\itemsep}{0pt}
  \item $l_{I_0}=1$    // input value, integer      
  \item $l_{c_1} = 2 $ // boolean constant False
  \item $l_{c_2}  = 3$ // integer constant 3
  \item $l_r     = 5$   // expected output value, boolean
  \item $l_{r_{f_1}} \in [4,5]$ // output of $f_1$,  boolean    
  \item $l_{r_{f_2}} \in [4,5]$ // output of $f_2$, boolean    
  \item $l_{I_{f_1,1}} \in \{l_{r_{False}}, l_{r_{f_1}}, l_{r_{f_2}}\}$   // param. of $f_1$, boolean
  \item $l_{I_{f_2,1}}\in \{l_{I_0}, l_{r_3}\}$   // first param. of $f_2$, integer
  \item $l_{I_{f_2,2}} \in \{l_{I_0}, l_{r_3}\}$  // second param. of $f_2$, integer
\end{description}

The following additional constraints are used to control the values of the location variables.
First, we need to make sure that there is only one building block output per building block input (wires are one-to-one).
$$\phi_{CONS}(L,O) = \bigwedge_{x,y \in O, x \neq y} l_x \neq l_y$$

Second, we need to order the building blocks in such a way that its arguments have already been defined.
$$\phi_{ACYC}(B,L,I,O,B) = \bigwedge_{(\phi_{i},I_i,r_i) \in B} \bigwedge_{x \in I_i} l_x < l_{r_i}$$
 
Putting all together:\\
$\phi_{WFF}(B,L,I,O, I_0,r) = \phi_{FIXED}(I_0,r) \land$ $\phi_{OUTPUT}(B)$\\ $ \land \phi_{INPUT}(I)$ $ \land \phi_{CONS}(L,I) \land \phi_{ACYC}(B,L,I,O)$\\

An assignment of $L$ variables respecting the predicate $\phi_{WFF}(B,L,I,O,I_0,r)$ corresponds to a syntactically correct patch. 

Values variables corresponding to the input and output of a building block are related according to its functional definition using a predicate $pb_i$ such that $pb_i(values(I_i),v_{r_i})=true$ iff $b_i(I_i)=r_i$. Let $V_{IO} = \{v_x | x \in I \cup O\}$. 

$$\phi_{LIB}(B,V_{IO}) = \bigwedge_{(I_i, r_i) \in B, v_{r_i}\in V_{IO}}\ pb_i(V_{IO}(I_i),v_{r_i})$$
The location and the value variables are connected together using the following rule which states that elements at the same location should have the same value. Note that because in our case
the input or output values can be of different types, we need to limit the application of that rule to
values of the same type. That limitation to the elements of the same type is valid because the
domain of the locations are managed using constraints $ \phi_{INPUT}(I)$.\\

$\phi_{CONN}(L,V_{IO}) =$ \\
$\bigwedge_{S \in \{BOOL,INT,REAL\}}\ \wedge_{x,y \in S}\ l_x = l_y \rightarrow v_x = v_y$

The semantic of the patch for a given input $I_0$ and a given output $r$ is preserved using the following existentially quantified constraint:\\

$\phi_{FUNC}(L,C_{l,m,n},O_{l,m,n}) =$ \\
$\exists V_{IO} \phi_{CONN}(L,V_{IO})[values(I_0) \leftarrow C_{l,m,n}, v_r \leftarrow O_{l,m,n}] $\\
$\land \phi_{LIB}(B,V_{IO})$

Here the notation $\alpha[v_r \leftarrow O_{l,m,n}]$ means that the value of the variable $v_r$ in $\alpha$ has been set to $O_{l,m,n}$.

Finally, finding a patch which satisfies all expected input/output pairs $(C_{l,m,n},O_{l,m,n})$ requires to satisfy the following constraint:\\
$\phi_{PATCH}(L,I,O,CO) =$\\
$ \exists L (\wedge_{(C_{l,m,n},O_{l,m,n}) \in CO}\ \phi_{FUNC}(L,C_{l,m,n},O_{l,m,n})) $\\
$\land \phi_{WFF}(B,L,I,O)$

\subsubsection{Levels}

Ideally, one would feed SMT with many instances of all kinds of building blocks we are considering (see \ref{sec:bb}). 
Only the required ones would be wired to the final result.
This is not efficient in practice. Some building blocks require computationally expensive theories (e.g. multiplication).
We first try to synthesize an expression with only one instance of easy building blocks ($<$, $\neq$, $=$, $\leq$\footnote{$\geq$ and $>$ are obtained by symmetry: $a~\geq~b = b~\leq~a$}).
Then, we add new building blocks (logic, then arithmetic) and eventually we increase the number of instances of building blocks.
This is encoded in arbitrary ``levels'' as Semfix does \cite{nguyen13}. The optimization of those predefined levels is future work.

\subsection{Deriving a Patch from an SMT model}

If the problem is satisfiable, the SMT solver provides an assignment to all location variables. Here is a possible answer for our running example:
$l_{i_0}   = 1$,              
$l_{c_1}   = 2    $,    
$l_{c_2}   = 3        $,
$l_{r_{final}}    = 5       $,
$l_{r_{f_1}}  = 4       $, 
$l_{r_{f_2}}  = 5       $, 
$l_{I_{f_1,1}}= 3  $,
$l_{I_{f_2,1}} = 1    $,
$l_{I_{f_2,2}} = 1  $.

The corresponding source patch is obtained with a backward traversal starting at the output location. There often exists building blocks which are wired (SMT produces a line number), but that are not connected to the final output of the expression.

In our example, this reads that the output is bound to line 5 which is the output of $f_2$.
$f_2$ takes as parameter line 1 which is the integer input value $i_0$.
The final patch is thus the expression $f_2(i_0, i_0)$ which returns a boolean. It is the repair of the bug, i.e. the fixed \ourif condition (or the missing precondition). In this example, $f_1$ is never used.

\section{Evaluation}
\label{sec:evaluation}

\nopol\ focuses on repairing conditional bugs in Java. In this section, we evaluate our approach with three case studies. First, we repair the running example of \cite{nguyen13}; then, we repair a real-world conditional bug from the Apache Commons Math library; finally, we show how to repair a missing precondition bug on an artificial example. 

Our prototype implementation of \nopol\ uses the Spoon library \cite{spoon} for manipulating Java source code (angelic value mining, instrumentation, final patch synthesis and assessment) and the GZoltar fault localization to order repair locations \cite{CamposRPA12}. \nopol\ generates SMTLIB files using jSMTLIB \cite{cok2011jsmtlib} and we use CVC4 \cite{CVC4} as SMT solver.
Thanks to the generic file format, \nopol\ can be used with any SMTLIB 2.0 compliant SMT solver.

\subsection{Case Study: Tcas Example from SemFix}
\label{case:semfix}
We first take a classical program, Tcas, which was used as example in previous work (SemFix \cite{nguyen13}). Tcas, a traffic collision avoidance system\footnote{Available in http://sir.unl.edu/}, is a program consisting of 135 lines of code, which originates from Software-artifact Infrastructure Repository (SIR) \cite{do2005supporting}. Figure \ref{fig:code-tcas} shows a code snippet of Tcas. For this code snippet, five test cases are listed in Table \ref{tab:test-case-tcas}. 

\begin{figure}[htbp]
\begin{verbatim}
1  int is_upward_preferred(boolean inhibit, 
     int up_sep, int down_sep) {
2     int bias;
3     if(inhibit)
4         bias = down_sep;   //fix: bias=up_sep+100
5     else
6         bias = up_sep;
7     if (bias > down_sep)
8         return 1;
9     else
10        return 0;
11   }
\end{verbatim}
\caption{The buggy method ``is\_upward\_preferred'' of Tcas}
\label{fig:code-tcas}
\end{figure}

\begin{table}[htbp]
\small
\setlength{\tabcolsep}{3pt}
\centering
\caption{Test suite with five test cases for Tcas}
\label{tab:test-case-tcas}
\begin{tabular}{c|c|c|c|c|c|c}

\hline
Test & \multicolumn{3}{|c|}{Input} & Expected & Observed & \multirow{2}{*}{Status} \\ 
\cline{2-4}
case & inhibit & up\_set & down\_set  &  output&  output& \\
\hline 
 1 & true & 0 & 100 & 0 & 0 & pass\\
 2 & true & 11 & 110 & 1 & 0 & fail \\
 3 & false & 100 & 50 & 1 & 1 & pass \\
 4 & true & -20 & 60 & 1 & 0 & fail \\ 
 5 & false & 0 & 10 & 0 & 0 & pass \\ 
\hline
\end{tabular}
\end{table}

As shown in Figure \ref{fig:code-tcas}, the faulty line in this code snippet is Line 4, where $down\_sep$ should be fixed with $up\_sep+100$. To our own surprise, \nopol\ generated a radically different patch as follows. For this faulty code, \nopol\ first employs angelic fix localization to find candidate lines. In this snippet, both Line 3 and Line 7 have the same likelihood of fault locations. 
However, angelic fix localization  states that only Line 7 is subject to repair (there exists angelic values for which the failing test cases pass), \nopol\ then collects testing traces by running the whole test suite. Finally, the testing trace is encoded as an SMT problem and \nopol\ outputs the following patch as solution to the bug: 

\begin{verbatim}
-      if (bias > down_sep)
+      if (up_sep != 0)
\end{verbatim}

To encode as an SMT program, the set of building blocks for Line 7 are list as follows. The input and output are $I_0=\{inhibit  \mbox{:BOOL}, up\_sep  \mbox{:INT}, down\_sep  \mbox{:INT}\}$ and $r=True$. 

\begin{description}
\setlength{\itemsep}{0pt}
\begin{scriptsize}
  \item $(f_{1} \mbox{:} I_{f_{1},1} \ <\  I_{f_{1},2}, I_{f_{1}}=\{I_{f_{1},1} \mbox{:INT} ,I_{f_{1},2} \mbox{:INT} \}, r_{f_{1}} \mbox{:BOOL} )$
  \item $(f_{2} \mbox{:} I_{f_{2},1} \ \leq\  I_{f_{2},2}, I_{f_{2}}=\{I_{f_{2},1} \mbox{:INT} ,I_{f_{2},2} \mbox{:INT} \}, r_{f_{2}} \mbox{:BOOL} )$
  \item $(f_{3} \mbox{:} I_{f_{3},1} == I_{f_{3},2}, I_{f_{3}}=\{I_{f_{3},1} \mbox{:INT} ,I_{f_{3},2} \mbox{:INT} \}, r_{f_{3}} \mbox{:BOOL} )$
  \item $(f_{4} \mbox{:} I_{f_{4},1} \ \neq\  I_{f_{4},2}, I_{f_{4}}=\{I_{f_{4},1} \mbox{:INT} ,I_{f_{4},2} \mbox{:INT} \}, r_{f_{4}} \mbox{:BOOL} )$
\end{scriptsize}
\end{description}

Among these four building blocks, comparison operators in Java are defined from $f_{1}$ to $f_{4}$, e.g., $f_{1}$ and $f_{2}$ denote $<$ and $\leq$, respectively.

After solving with SMT, the solution is $f_{4}(up\_sep, 0))$, i.e., up\_sep != 0 as mentioned above. This patch is not identical to the fix discussed in the SemFix paper \cite{nguyen13} (shown as comment in Figure \ref{fig:code-tcas} at line 7). 
However, from the perspective of test-suite based program repair, \nopol\ patch is also a correct patch since it passes all test cases. The reason for the existence of two different patches is that the test suite in Table \ref{tab:test-case-tcas} does not provide test cases to differentiate the two patches.    

\subsection{Case Study: Commons-Math Library}
\label{sec:bug-math}

In this section, we present how \nopol\ is able to repair a real bug\footnote{See details, https://github.com/apache/commons-math/commit/23277b96dae68928fbe7a785d080fc0fbf6eb27d} in Apache Commons Math.  Apache Commons Math is a lightweight library for common mathematics and statistics problems\footnote{http://commons.apache.org/proper/commons-math/}. This library consists of 5000 lines of executable code and 352 test cases (each test case is expressed as a JUnit method). Figure \ref{fig:code-math} shows the buggy source code of its Percentile class. 
As its name suggests, Percentile returns an estimate of the $p$th percentile of the values stored in the array $values$. 

\begin{figure}[htbp]
\begin{verbatim}
1  public double evaluate(final double[] values,
     final double p){
     ...
2    double n = values.length;
     ...
3    double pos = p * (n + 1) / 100;
4    double fpos = Math.floor(pos);
5    int intPos = (int) fpos;
6    double dif = pos - fpos;
7    double[] sorted = new double[n];
8    System.arraycopy(values, 0, sorted, 0, n);
9    Arrays.sort(sorted);
10   if (pos < 1) 
11       return sorted[0];   
12   if (pos > n)     //fix: if (pos >= n) 
13       return sorted[n - 1];    
14   double lower = sorted[intPos - 1];
15   double upper = sorted[intPos];
16   return lower + dif * (upper - lower);
17 }	
\end{verbatim}
\caption{Code snippet of Percentile in Commons Math}
\label{fig:code-math}
\end{figure}

According to the documentation, the algorithm of Percentile is implemented as follows. 
Let $n$ be the length of the (sorted) array. Compute the estimated percentile position $pos = p * (n + 1) / 100$ and the difference, $dif$ between $pos$ and $floor(pos)$. If $pos >= n$ return the largest element in the array; otherwise return the final calculation of percentile. Thus, Line 12 in Figure \ref{fig:code-math} contains a bug, which should be corrected as $if(pos >= n)$.  
Table \ref{tab:test-case-percentile} shows one failing test case exists for this bug and one of the 351 passing test cases. For the failing test case, an $ArrayIndexOutOfBounds$ exception is thrown in Line 15. 

\begin{table}[htbp]
\small
\centering
\caption{Test suite with one failed test case for Percentile}
\label{tab:test-case-percentile}
\begin{tabular}{c|c|c|c|c}

\hline
 \multicolumn{2}{c|}{Input} & \multicolumn{2}{|c|}{Output evaluate(values,p)} & \multirow{2}{*}{Status} \\ 
\cline{1-4}
 values & p &  Expected&  Observed& \\
\hline 
 \{0,1\} & 25 & 0.0 & 0.0 & pass\\
 \{1,2,3\} & 75 &  3.0 & Exception  & fail\\
\hline
\end{tabular}
\end{table}

The building blocks for this program are as follows (the same as building blocks in Section \ref{case:semfix}, except replacing INT with REAL).  

\begin{description}
\setlength{\itemsep}{0pt}
\begin{scriptsize}
  \item $(f_{1} \mbox{:} I_{f_{1},1} \ <\  I_{f_{1},2}, I_{f_{1}}=\{I_{f_{1},1} \mbox{:REAL} ,I_{f_{1},2} \mbox{:REAL} \}, r_{f_{1}} \mbox{:BOOL} )$
  \item $(f_{2} \mbox{:} I_{f_{2},1} \ \leq\  I_{f_{2},2}, I_{f_{2}}=\{I_{f_{2},1} \mbox{:REAL} ,I_{f_{2},2} \mbox{:REAL} \}, r_{f_{2}} \mbox{:BOOL} )$
  \item $(f_{3} \mbox{:} I_{f_{3},1} == I_{f_{3},2}, I_{f_{3}}=\{I_{f_{3},1} \mbox{:REAL} ,I_{f_{3},2} \mbox{:REAL} \}, r_{f_{3}} \mbox{:BOOL} )$
  \item $(f_{4} \mbox{:} I_{f_{4},1} \ \neq\  I_{f_{4},2}, I_{f_{4}}=\{I_{f_{4},1} \mbox{:REAL} ,I_{f_{4},2} \mbox{:REAL} \}, r_{f_{4}} \mbox{:BOOL} )$
\end{scriptsize}
\end{description}

Based on the above building blocks, the solution to SMT is $f_{2}(n, pos)$. That is pos >= n in Java. Thus, \nopol\ can generate the expected patch ($if(pos >= n)$). \nopol's angelic fix localization was able to spot a fixable \ourif condition after analyzing 11 candidate \ourif conditions.
In contrast with the example of Section \ref{case:semfix}, this is a real-world use case. Based on a test suite of 352 test cases with only one failed test case, \nopol\ finds a patch that matches the patch made by the developers. 

\subsection{Case Study: Missing Precondition Example}
\label{sec:bug-misspre}

In this section, we take an artificial example to show how to repair a missing-precondition bug with \nopol. Figure \ref{fig:code-missing} presents a method for extracting the folder of a given path. Its related test suite consists of two test cases: one for an absolute File path, in which case the method should return the enclosing folder; one for a local file, in which case an empty string is expected. The bug of this method is a missing precondition in Line 5.
In the absence of this precondition, an $ArrayIndexOutOfBounds$ exception is thrown in Line 6. 

\begin{figure}[htbp]
\begin{verbatim}
   //In Linux, separator is "/"
1  private final String separator = File.separator; 
2  public String extractFolder(String path)  {	
3      String result = "";
4      int index = path.lastIndexOf(separator);

5      //fix: if(index > 0)
6      result = path.substring(0, index);
7      return result;
8  }
\end{verbatim}
\caption{Code snippet of an example of missing-precondition bug}
\label{fig:code-missing}
\end{figure}

\begin{table}[htbp]
\setlength{\tabcolsep}{3pt}
\small
\centering
\caption{Test suite with two test case for a missing precondition bug}
\label{tab:test-case-missing}
\begin{tabular}{c|c|c|c|c}

\hline
Test & Input & \multicolumn{2}{|c|}{Output} & \multirow{2}{*}{Status} \\ 
\cline{3-4}
case & path &  Expected&  Observed& \\
\hline 
 1 & /home/user/Path.java & /home/user & /home/user & pass \\
 2 & File.java &  & Exception & fail \\
\hline
\end{tabular}
\end{table}

With \nopol, the bug in Line 5 can be repaired. Building blocks for Line 5 are the same as those in Section \ref{case:semfix}. Then the solution to the SMT based on these above building blocks is $f_{1}(result.length(),index)$. Thus, the result of \nopol\ is as follows.  
\begin{verbatim}
+  if(result.length() < index)
\end{verbatim}
Note that this repair is a little different from Line 5 in Figure \ref{fig:code-missing}. Since the length of $result$ is $0$, the repair by \nopol\ is equivalent to Line 5. For this bug, \nopol\ finds no candidate buggy \ourif conditions with angelic fix localization. 
When it studies whether the bug might be a missing precondition, it tries to skip all statements of the method.
After adding an angelic precondition at Line 6, the failing test passes. Then, the runtime data is encoded as SMT and the SMT solution is converted as the final patch.   

\section{Limitations}

In this section, we list several limitations of our work. 

\textbf{Single point of failure} As most of the previous work in test-suite based program repair, our approach, \nopol, can only deal with faulty programs, where there is only one fault. In the current version, we do not stack patches (the search space would be exponentially larger). 

\textbf{Granularity of building blocks} The ability to fix a bug heavily depends on the buildings blocks expressible in SMT. If $B$ is too small, there are few chances to fix the bug, unless it is trivial. If $B$ is too large, the generated patch may be too large and complex to be accepted by a developer since we have no warrantee of minimality. Improving the way we set the building blocks of $B$ is in our research agenda. 

\textbf{Non-fixable bugs}
\nopol\ cannot generate patches for all buggy \ourif conditions and missing preconditions. The ability of patch generation is limited by the  test suite. In our work, for a buggy \ourif condition, we use angelic fix localization to flip the boolean value of the condition for the failing test cases. However, if no passing test case covers this flipped boolean value, a trivial repair would be to replace the conditional expression by a boolean constant. It means that to generate a non trivial patch for a buggy \ourif condition, both boolean values of a condition should be covered by at least one passing and one failing test cases. 

\textbf{Conditions including methods with parameters}
In our work, we can currently support the synthesis of conditionals with methods without parameters.
Our approach cannot generate a patch if a method with parameters has to appear in a condition. For example, no patch can be generated for $if(list.contains(object))$ due to its parameter $object$. The reason is that 
since we need to gather the possible values of the method call, we would need to evaluate that method for each possible parameter when running the tests. While it is feasible for methods with one parameter, methods with more parameters would induce a combinatoric step during the data collection phase. We plan to support methods with few parameters in the future by leveraging semantic analysis. 

\textbf{Limitations of SMT solvers}
In the current state of our research, the performance and intrinsic limitations of SMT is not an issue. 
Given our set of constraints (as given by our encoding of the repair problem), the solver we use (CVC4) is sufficient for our needs (it takes a few seconds to fix the Apache Common Maths bug). However, our evaluation is preliminary. In particular, we anticipate one major  problem: CVC4 does not have currently a full support for non linear arithmetic. This is problematic for synthesizing conditional expressions that use multiplication or division. We consider using another SMT solver (e.g Z3) in order to overcome this issue.

\section{Related Work}

Test-suite based program repair generates patches and examines patches with a given test suite. Le~Goues et al. \cite{Goues2012} propose GenProg, a test-suite based program repair approach using genetic programming. In GenProg, a program is viewed as an Abstract Syntax Tree (AST) while a patch is a newly-generated AST by weighting statements in the program. Based on genetic programming, candidate patches are generated via multiple trials. Then for each candidate patch, a given test suite is evaluated to identify the patch with all passing test cases. The role of genetic programming is to obtain new ASTs by copying and replacing nodes in the original AST. An evaluation on 16 C programs shows that GenProg can generate patches with an average success rate of 77 percent. 

Nguyen et al. \cite{nguyen13} propose SemFix, a program repair approach via semantic analysis. In contrast to genetic programming in GenProg, SemFix generates patches by combining symbolic execution, constraint solving, and program synthesis. As mentioned in Section \ref{sec:oracle-based-repair}, SemFix generate constraints by formulating passed test cases and then solve such constraints via traversing a search space of repair expressions. Compared with GenProg, SemFix reports a higher success rate on C programs within less time cost. In this paper, we also focus on program repair by leveraging constraint solving and program synthesis. The key difference compared to SemFix is that \nopol is able to repair missing preconditions, a kind of fault that is not handled by SemFix.  

Kim et al. \cite{Kim2013} propose Par, a repair approach using fix patterns representing common ways of fixing common bugs. These fix patterns can avoid the nonsensical patches due to the randomness of some mutation operators.  Based on the fix patterns, 119 bugs are examined for patch generation. In this work, the evaluation of patches are contributed by 253 human subjects, including 89 students and 164 developers. 

Martinez and Monperrus \cite{Martinez2013} mine historical repair actions to reasoning future actions with a probabilistic model. Based on a fine granularity of abstract syntax trees, this work analyzes over 62 thousands versioning transactions in 14 repositories of open-source Java projects to collect probabilistic distributions of repair actions. Such distributions can be used as prior knowledge to guide program repairing. 
 
Program synthesis is a related topic to program repair. Program synthesis aims to form a new program by synthesizing existing program components. Jha et al. \cite{jha2010oracle} mine program oracles based on examples and employs SMT solvers to synthesize constraints. In this work, manual or formal specifications are replaced by input-output oracles. They evaluate this work on 25 benchmark examples in program deobfuscation. Their following work \cite{gulwani2011synthesis} addresses the same problem by encoding the synthesis constraint with a first-order logic formula. The maximum size of constraint is quadratic in the number of given components.    

In our work, fault localization is used as a step to provide faulty statements. The goal of fault localization \cite{do2005supporting} is to rank suspicious statements (or blocks, classes) to find out the location of bugs. A general framework of fault localization is to collect program spectrum (a matrix of testing results based on a given test suite) and to sort statements in the spectrum with specific metrics (e.g., Tarantula \cite{do2005supporting} and Ochiai \cite{abreu2006evaluation}). Among existing metrics in fault localization, Ochiai \cite{abreu2006evaluation} has been evaluated as one of the most effective ones. In Ochiai, each statement is assigned with its suspiciousness value, which is the Ochiai index between the number of failed test cases and the number of covered test cases. 

\section{Conclusion}

In this paper, we propose \nopol, a test-suite based repair approach using SMT. We target two kinds of bugs: buggy \ourif conditions and missing preconditions. Given a faulty program and its test suite, \nopol\ employs a specific fault localization technique, angelic fix localization, to find suspicious statements. For each candidate statement, \nopol\ collects test execution traces at this point of the program. Those traces are then encoded as an SMT problem and the solution to this SMT is converted into a patch for the faulty program. Preliminary results on a real-world bug in Apache Common Maths library and two artificial examples show that our approach can fix the bugs of our fault model: buggy \ourif conditions and missing preconditions. 

In future work, we plan to evaluate our approach on more real-world bugs. We also would like to extend \nopol\ to fix bugs in conditionals of loop structures (while, for, etc.). 

\section*{Acknowledgments}
This work is partially supported by the INRIA Internships program and the CNRS delegation program.
We would like to thank David Cok for giving us full access to jSMTLIB.

\bibliographystyle{plain} 
\balance
\bibliography{vee99878a7c154b75}

\end{document}